\documentclass[aps,prl,twocolumn,superscriptaddress]{revtex4-1}

\usepackage{amsmath,amssymb,graphicx}
\usepackage{url}
\usepackage[colorlinks,urlcolor=blue]{hyperref}

\begin{document}

\title{3D Structure of Jet-Induced Diffusion Wake in an Expanding Quark-Gluon Plasma}

\author{Zhong Yang}
\affiliation{Key Laboratory of Quark and Lepton Physics (MOE) \& Institute of Particle Physics, Central China Normal University, Wuhan 430079, China}

\author{Tan Luo}
\email[]{tan.luo@usc.es}
\affiliation{Instituto Galego de F\'isica de Altas Enerx\'ias IGFAE, Universidade de Santiago de Compostela, E-15782 Galicia-Spain}
 
\author{Wei Chen}
\email[]{chenwei@ucas.ac.cn}
\affiliation{School of Nuclear Science and Technology, University of Chinese Academy of Sciences, Beijing 100049, China}

\author{Longgang Pang}
\email[]{lgpang@mail.ccnu.edu.cn}
\affiliation{Key Laboratory of Quark and Lepton Physics (MOE) \& Institute of Particle Physics, Central China Normal University, Wuhan 430079, China}

\author{Xin-Nian Wang}
\email[]{xnwang@lbl.gov}
\affiliation{Key Laboratory of Quark and Lepton Physics (MOE) \& Institute of Particle Physics, Central China Normal University, Wuhan 430079, China}
\affiliation{Nuclear Science Division MS 70R0319, Lawrence Berkeley National Laboratory, Berkeley, California 94720, USA}
\thanks{Current address}

\begin{abstract}
The diffusion wake accompanying the jet-induced Mach cone provides a unique probe of the properties of quark-gluon plasma in high-energy heavy-ion collisions. It can be characterized by a depletion of soft hadrons in the opposite direction of the propagating jet.
We explore the 3D structure of the diffusion wake induced by $\gamma$-triggered jets in Pb+Pb collisions at the LHC energy within the coupled linear Boltzmann transport and hydro model. We identify a valley structure caused by the diffusion wake on top of a ridge from the initial multiple parton interaction (MPI) in jet-hadron correlation as a function of rapidity and azimuthal angle. This leads to a double-peak structure in the rapidity distribution of soft hadrons in the opposite direction of the jets as an unambiguous signal of the diffusion wake. Using a two-Gaussian fit, we extract the diffusion wake and MPI contributions to the double peak. The diffusion wake valley is found to deepen with the jet energy loss as characterized by the $\gamma$-jet asymmetry. Its sensitivity to the equation of state and shear viscosity is also studied.
\end{abstract}

\maketitle

\noindent{\it \color{blue} 1. Introduction.---}Projectiles traveling at a speed faster than the velocity of sound generate Mach waves in the medium which can combine to become a shock wave known as the Mach cone \cite{mach}, such as the sonic boom originating from a supersonic aircraft. These kinds of Mach cones are also produced on the femtometer scale inside the quark-gluon plasma (QGP) by propagating jets in ultrarelativistic heavy-ion collisions \cite{Casalderrey-Solana:2004fdk, Stoecker:2004qu, Ruppert:2005uz, Gubser:2007ga, Qin:2009uh,Li:2010ts,Bouras:2012mh,Ayala:2016pvm,Yan:2017rku,Cao:2020wlm}. Study of such jet-induced Mach cones can provide important information about the properties of the QGP such as transport coefficients and the equation of state (EOS).

Though the spatial structure of jet-induced Mach cones in heavy-ion collisions is very distinctive according to the full relativistic fluid dynamics \cite{Qin:2009uh,Yan:2017rku} and linearized hydrodynamic studies \cite{Neufeld:2008fi,Casalderrey-Solana:2020rsj}, its signal in the final hadron spectra \cite{Li:2010ts,Bouras:2012mh, Ayala:2016pvm,Renk:2007rv,Renk:2006mv,Ma:2010dv,Betz:2010qh,Tachibana:2020mtb,Tachibana:2014lja,Tachibana:2017syd,Luo:2021voy} has never been unambiguously observed. The complication is caused by soft gluon radiations from the propagating jet and the average over the propagation path and direction.  For example, soft hadrons from the jet-induced Mach cone have a distinctive azimuthal angle distribution when the jet propagation path and angle are fixed relative to the radial flow of the expanding QGP \cite{Li:2010ts,Tachibana:2020mtb}. However, the final distribution after averaging over all possible angles and path lengths becomes a nondescript Gaussian similar to that of soft hadrons from jet fragmentation \cite{Chen:2017zte,Yang:2021qtl}. The energy scale $\omega \sim T$ for radiative gluons induced by jet-medium interaction \cite{Blaizot:2013hx,Schlichting:2020lef, Ke:2020clc} is also similar to that from the jet-induced Mach cone.

Similar to any type of shock waves generated by fast projectiles, jet-induced Mach cones in QGP are always accompanied by a diffusion wake behind the propagating jet as a general phenomenon in the hydrodynamic description of the medium response to the energy-momentum deposition \cite{Neufeld:2008dx,Betz:2008ka}. Microscopically in a transport description, jet-medium interaction kicks the medium parton into a recoil particle and leaves behind a particle-hole. Further transport of these recoil particles forms the Mach waves, while the diffusion of the particle holes leads to the diffusion wake \cite{He:2015pra}. Such diffusion wakes will lead to a depletion of soft hadrons in the final hadron spectra in the opposite direction of the jets \cite{Chaudhuri:2006qk,Chaudhuri:2007vc,Tachibana:2015qxa,Chen:2017zte,Pablos:2019ngg,Yang:2021qtl,Yan:2017rku} which can be considered as an unambiguous signal of the medium response accompanying the jet-induced Mach cone. Medium modification of partons from the initial multiple parton interaction (MPI), however, is found to give rise to a uniform (in azimuthal angle) soft hadron enhancement that can overwhelm the deletion due to the diffusion wake \cite{Yang:2021qtl}. One therefore needs to separate the contribution from MPI with a mixed-event procedure in order to observe the diffusion wake. One can further use a 2D jet tomography \cite{He:2020iow} to select events with longer jet propagation lengths to enhance the signal of the jet-induced diffusion wake.

In this Letter, we explore the 3D structure of the jet-induced diffusion wake in $\gamma$-jet events in Pb+Pb collisions at the Large Hadron Collider (LHC) within the coupled linear Boltzmann transport and hydro (CoLBT-hydro) model. We will examine in particular the jet-hadron correlation in rapidity and azimuthal angle. We will show that the diffusion wake leads to a unique valley structure in the opposite direction of the $\gamma$-triggered jet on top of a ridge from MPI. Using a two-Gaussian fit we extract the MPI ridge and the diffusion wake valley. The diffusion wake valley is shown to deepen with the jet energy loss as characterized by the $\gamma$-jet asymmetry while MPI ridge remains approximately the same. We will also study the sensitivity of the diffusion wake valley to EOS and shear viscosity.

\noindent{\em \color{blue} 2. Jet quenching within the CoLBT-hydro model.---}
Jet-induced Mach cone and the diffusion wake arise from the propagation of recoil partons and the diffusion of particle holes in a microscopic transport picture. Macroscopically, they can also be described by the collective response from the energy-momentum deposited into the QGP by the propagating jet in a hydrodynamic approach.
In this study, we use the CoLBT-hydro model \cite{Chen:2017zte,Chen:2020tbl,Zhao:2021vmu} to simulate (direct)$\gamma$-jet propagation and jet-induced medium response in Pb+Pb collisions at the LHC. CoLBT-hydro combines the microscopic linear Boltzmann transport (LBT) model \cite{He:2015pra} for the propagation of energetic jets and recoil partons with the event-by-event (3+1)D CCNU-LBNL viscous (CLVisc) hydrodynamic model \cite{Pang:2012he,Pang:2014ipa,Pang:2018zzo} for the evolution of the bulk medium and soft modes of the jet-induced medium response. LBT and CLVisc are coupled in real time through a source term from the energy momentum lost to the medium by jet shower and recoil partons as well as the particle holes or ``negative partons" from the backreaction. The LBT model \cite{He:2015pra} is based on the Boltzmann equation for both jet shower and recoil partons with perturbative QCD (pQCD) leading-order elastic scattering and induced gluon radiation according to the high-twist approach \cite{Guo:2000nz,Wang:2001ifa,Zhang:2003yn,Zhang:2003wk}. CLVisc~\cite{Pang:2014ipa,Pang:2012he,Pang:2018zzo} parallelizes the Kurganov-Tadmor algorithm \cite{KURGANOV2000241} to solve the hydrodynamic equation for the bulk medium including medium response and Cooper-Frye particlization on GPU. A freeze-out temperature $T_f=137$ MeV and specific shear viscosity $\eta/s=0.15$ together with the {\it s95p} parameterization of the lattice QCD EOS with a rapid crossover \cite{Huovinen:2009yb} and Trento \cite{Moreland:2014oya} initial conditions with a longitudinal envelope at an initial time $\tau_0=0.6$ fm/$c$ are used to reproduce experimental data on bulk hadron spectra and anisotropic flows at the LHC energies \cite{Pang:2018zzo}. The Trento model is also used to provide the transverse spatial distribution of $\gamma$-jets whose initial configurations are generated from PYTHIA8 \cite{Sjostrand:2007gs}. 
Partons from the initial jet showers as well as MPI propagate through the QGP and generate medium response according to the CoLBT-hydro model. The final hadron spectra include contributions from the hadronization of hard partons within a parton recombination model \cite{Han:2016uhh,Zhao:2020wcd} and jet-induced hydro response via Cooper-Frye freeze-out after subtracting the background from the same hydro event without the $\gamma$-jet.  For detailed descriptions of the LBT and CoLBT-hydro model we refer readers to Refs.~\cite{He:2015pra,Cao:2016gvr,Cao:2017hhk,He:2018xjv,Luo:2018pto,Zhang:2018urd} and \cite{Chen:2017zte,Zhao:2021vmu,Yang:2021qtl}.

\begin{figure}
\centerline{\includegraphics[scale=0.7]{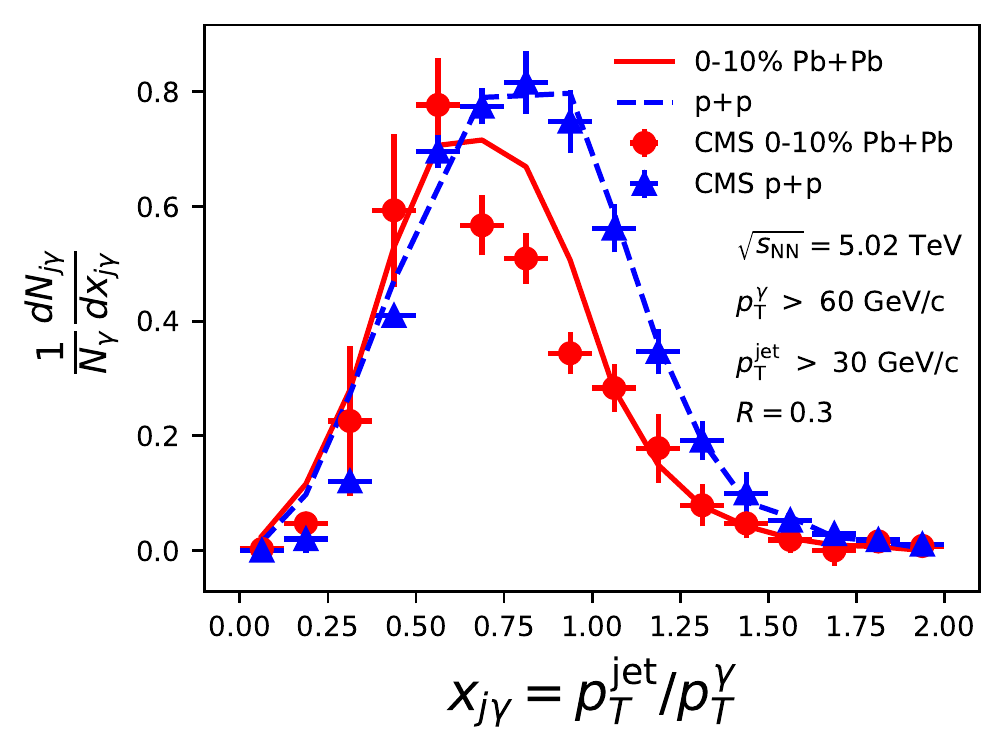}}
\caption{Event distribution in $\gamma$-jet asymmetry $x_{j\gamma}=p_{\rm T}^{\rm jet}/p_{\rm T}^\gamma$ in $p+p$ (blue dashed) and Pb+Pb (red solid) collisions at $\sqrt{s_{\rm NN}}=5.02$ TeV from CoLBT-hydro as compared to CMS data \cite{CMS:2017ehl}, for $p_{\rm T}^{\gamma}>60$ GeV/$c$,  $p_{\rm T}^{\rm jet}>30$ GeV/$c$, $\left |\eta^\gamma \right |<$1.44 and $\left | \eta^{\rm jet} \right | <$1.6, $\left | \phi_\gamma - \phi_{\rm jet} \right |>$ $7/8\pi$ and jet cone size $R=0.3$.}
\label{fig:asym}
\end{figure}

The energy loss by propagating jet shower partons that generate the Mach cone and diffusion wake  also leads to modification of the final reconstructed jets \cite{He:2018xjv,Luo:2018pto,Zhang:2018urd,Casalderrey-Solana:2015vaa, Casalderrey-Solana:2016jvj,Chen:2020tbl}. Shown in Fig.~\ref{fig:asym} are the $\gamma$-jet asymmetry distributions from CoLBT-hydro in p+p (dashed) and 0-10\% central Pb+Pb (solid) collisions at $\sqrt{s_{\rm NN}}=5.02$ TeV as compared to CMS data~\cite{CMS:2017ehl}. FASTJET~\cite{Cacciari:2011ma} is used to reconstruct jets with the anti-$k_{\rm T}$ algorithm and jet cone size $\rm R$=0.3. The same smearing due to jet-energy resolution is applied to CoLBT-hydro results as in the CMS data. Since photons do not interact with the QGP, jet energy loss in Pb+Pb collisions will lead to a smaller value of the $\gamma$-jet asymmetry $x_{j\gamma}=p_{\rm T}^{\rm jet}/p_{\rm T}^{\gamma}$ as compared to $p+p$ collisions as shown by both the experimental data and CoLBT-hydro with an effective strong coupling constant $\alpha_s=$ 0.24. As we will show later in this Letter, the $\gamma$-jet asymmetry $x_{j\gamma}$ can be used to control the average jet energy loss and consequently the jet-induced diffusion wake. 

\begin{figure}
\centerline{\includegraphics[scale=0.85]{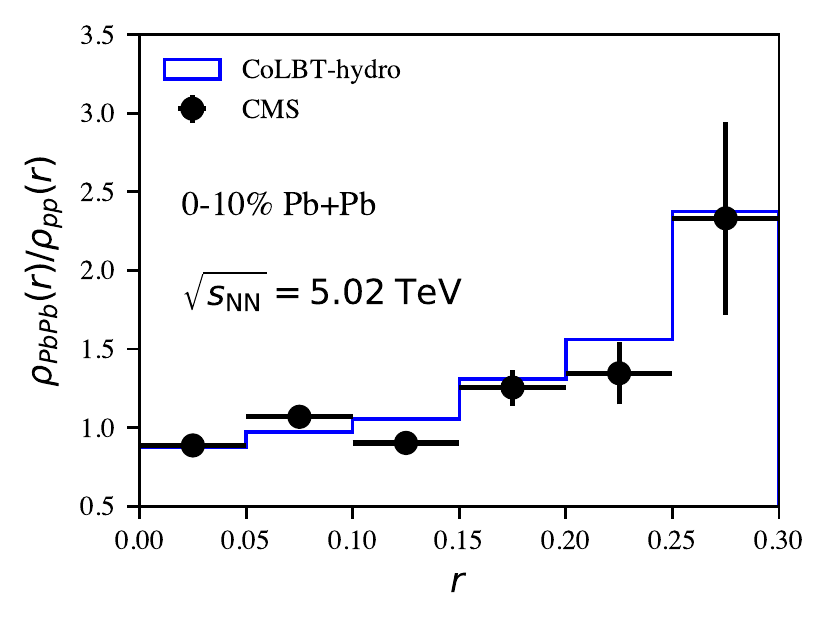}}
\caption{Ratio of $\gamma$-jet shape in Pb+Pb over that in $p+p$ collisions at $\sqrt{s_{\rm NN}}=5.02$ TeV from CoLBT-hydro as compared to the CMS data \cite{CMS:2018jco}, for the same kinematics as in Fig.~\ref{fig:asym}.}
\label{fig:shape}
\end{figure}

To check the influence of the jet-induced medium response on the final jet structure, we show in Fig.~\ref{fig:shape} the modification of the jet shape,
\begin{equation}
\rho(r)=\frac{1}{\delta r}\frac{\sum_{\rm jets}\sum_{r<r_{\rm trk}<r+\delta r}(p_{\rm T}^{\rm trk}/p_{\rm T}^{\rm jet})}{\sum_{\rm jets}\sum_{r_{\rm trk}<R}(p_{\rm T}^{\rm trk}/p_{\rm T}^{\rm jet})},
\end{equation}
where $p_{\rm T}^{\rm trk}>1$ GeV/$c$ is the transverse momentum of a charged track, $r=\sqrt{(\phi^{\rm trk}-\phi^{\rm jet})^2+(\eta^{\rm trk} -\phi^{\rm jet})^2}$ is the distance between the track and the jet axis in rapidity ($\eta$) and azimuthal angle ($\phi$). The summation is over all jets within the kinematic cuts and over an annulus of width $\delta r$ with respect to the jet axis in the numerator and over the jet cone $R=0.3$ in the denominator. Both CoLBT-hydro and experimental data \cite{CMS:2018jco} show a significant broadening of the jet shape toward the edge of the jet cone due to the jet-induced medium response and medium-induced gluon radiation. The same mechanisms also lead to the enhancement of soft hadrons in the jet fragmentation function \cite{Chen:2017zte,Chen:2020tbl}.  However, it is difficult to separate the two mechanisms in both the experimental data and CoLBT-hydro simulations.

\begin{figure}
\centerline{\includegraphics[scale=0.7]{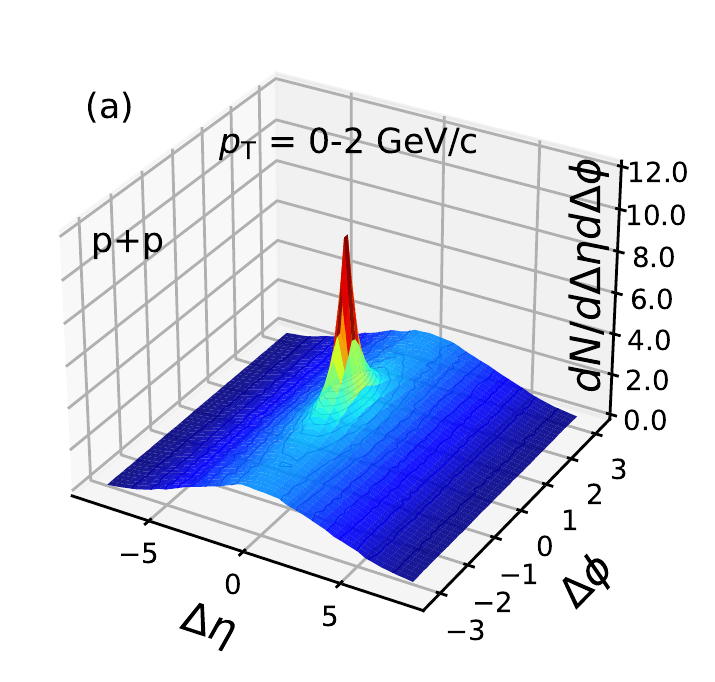}
\hspace{-0.18in}\includegraphics[scale=0.7]{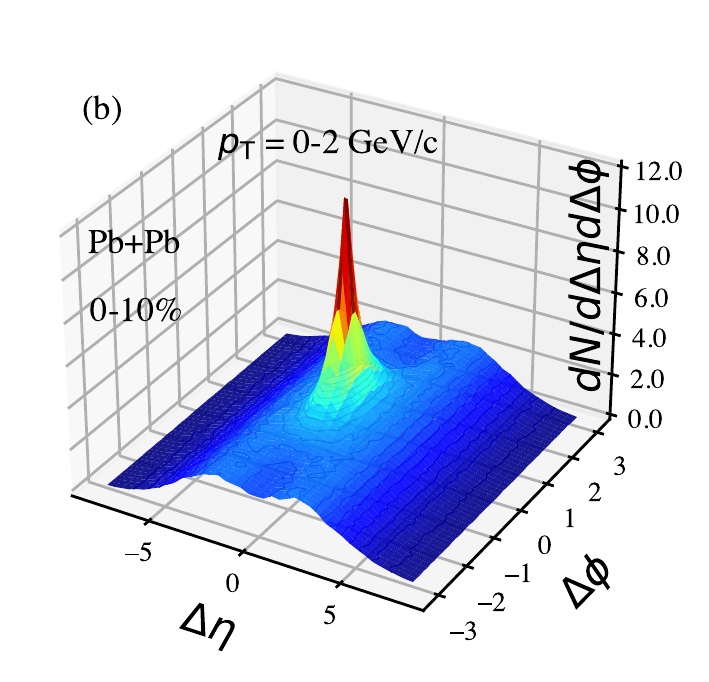}}
\caption{CoLBT-hydro results on $\gamma$-triggered jet-hadron correlation for soft hadrons ($p_{\rm T}=0$-2 GeV/$c$) in $\Delta\eta=\eta_h-\eta_{\rm jet}$ and $\Delta\phi=\phi_h-\phi_{\rm jet}$ in (a) $p+p$ and (b) 0-10\% Pb+Pb collisions at $\sqrt{s_{\rm NN}}=5.02$ TeV, with the same kinematics as in Fig.~\ref{fig:asym}.}
\label{fig:ghadron}
\end{figure}

\noindent{\em \color{blue} 3. 3D structure of the diffusion wake.---}To find out the 3D structure of the jet-induced medium response in the momentum space, we plot in Fig.~\ref{fig:ghadron} the jet-hadron correlations in $\Delta \eta=\eta_h-\eta_{\rm jet}$ and $\Delta \phi=\phi_h-\phi_{\rm jet}$ for soft hadrons in $p_{\rm T} \in (0,2)$ GeV/$c$ in (a) $p+p$ and (b) 0-10\% central Pb+Pb collisions at $\sqrt{s_{\rm NN}}=5.02$ TeV.  The correlation in $p+p$ collisions has a peak around the jet axis for hadrons from the jet on top of a ridge along the azimuthal angle  from MPI (a small fraction $\sim 20\%$ of this ridge comes from initial state radiation).  In Pb+Pb collisions, the jet peak is clearly enhanced by both the recoil and radiated partons as a result of the jet modification. This is consistent with soft hadron enhancement in the modified jet fragmentation functions \cite{Chen:2017zte,Chen:2020tbl,Yang:2021qtl}. In the azimuthal angle region $|\Delta\phi|>\pi/2$ opposite to the jet axis around $|\Delta\phi|=\pi$, however, a valley is formed on top of the MPI ridge due to the depletion of soft hadrons by jet-induced diffusion wake.
We refer this as the diffusion wake (DF-wake) valley. We will focus on the structure of this valley in rapidity $\Delta \eta$ as a unique signal of the diffusion wake in the remainder of this Letter.

\begin{figure}
\centerline{\includegraphics[scale=0.5]{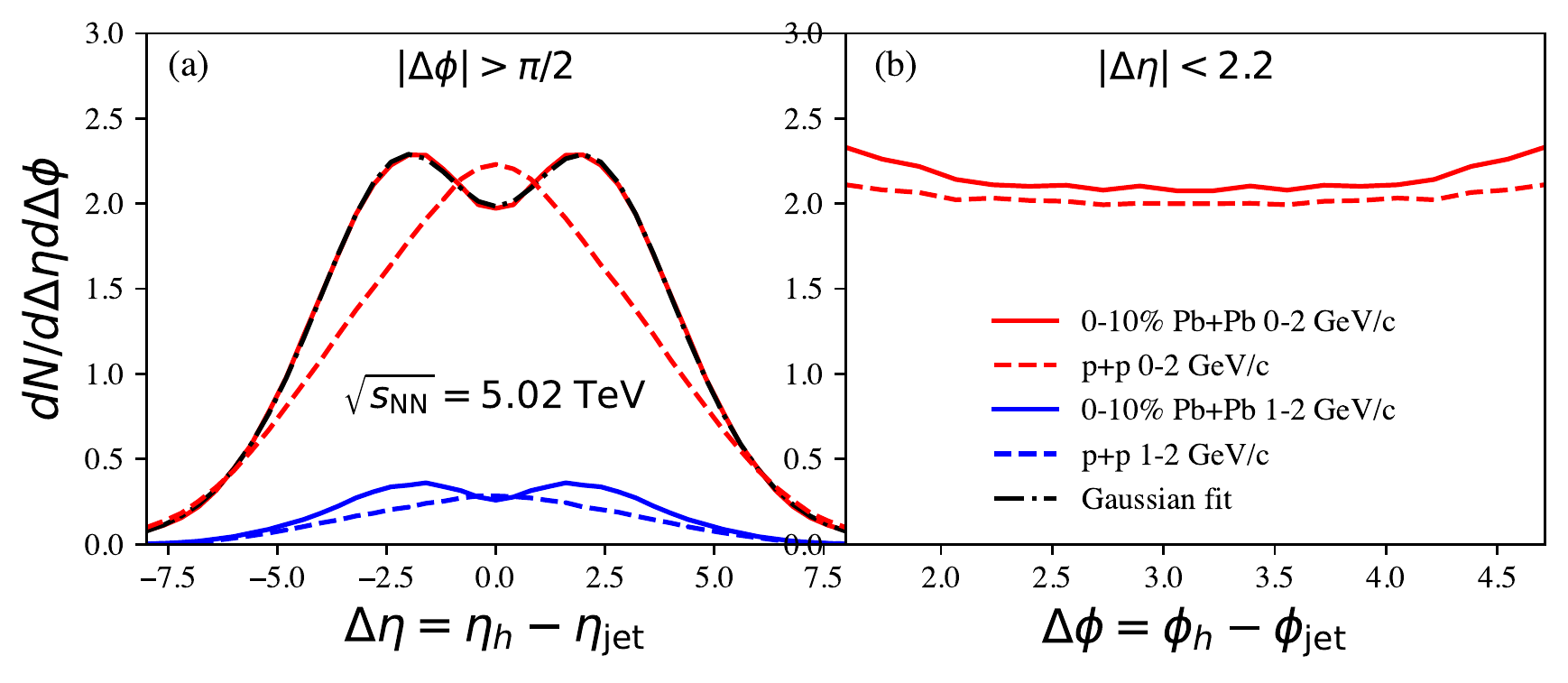}}
\caption{CoLBT-hydro results on $\gamma$-triggered jet-hadron correlation (a) in $\Delta\eta$ within $|\Delta\phi|>\pi/2$ and (b) in $\Delta\phi$ within $|\Delta\eta|<2.2$ for soft hadrons within $p_{\rm T}=0$-2 GeV/$c$ (red) and $p_{\rm T}=1$-2 GeV/$c$ range (blue) in $p+p$ (dashed)  and 0-10\% central Pb+Pb (solid) collisions at $\sqrt{s_{\rm NN}}=5.02$ TeV. The $\gamma$-jet kinematics is the same as in Fig.~\ref{fig:asym}. The black dot-dashed line is the 2-Gaussian fit using Eq.~(\ref{fit}).}
\label{fig:etacorr}
\end{figure}

To examine the structure of the DF-wake valley in detail, we plot in Fig.~\ref{fig:etacorr} the jet-hadron correlation (a) as a function of rapidity $\Delta \eta$ in the region  $|\Delta\phi|>\pi/2$ and (b) as a function of $\Delta\phi$ in the region  $|\Delta\eta|<2.2$ for soft hadrons in $p+p$ (dashed) and 0-10\% central Pb+Pb (solid) collisions \cite{footnote}.  The Gaussian-like MPI ridge of the correlation in $p+p$ collisions comes from independent minijets in MPI. In Pb+Pb collisions, these minijets are also quenched, leading to enhancement of soft hadrons and suppression of high $p_{\rm T}$ hadrons. Their rapidity-azimuthal distributions, however, remain a Gaussian-like ridge plus a valley due to the diffusion wake. The DF-wake valley on top of the MPI ridge gives rise to a double peak feature in the rapidity distribution of the jet-hadron correlation in Fig.~\ref{fig:etacorr} (a). The DF-wake valley is the deepest in the direction opposite  to the jet axis ($|\Delta\phi|=\pi$). As one moves toward the jet axis in azimuthal angle, the valley gradually gives away to the jet peak starting at around $|\Delta\phi|\leq\pi/2$ as seen in Figs.~\ref{fig:ghadron}(b) and \ref{fig:etacorr}(b).

In order to disentangle the DF-wake valley and MPI ridge in the jet-hadron correlation, we use a 2-Gaussian,
\begin{equation}
F(\Delta\eta)=\int_{\eta_{j1}}^{\eta_{j2}} d\eta_j F_3(\eta_j)(F_2(\Delta\eta,\eta_j)+F_1(\Delta\eta)),
\label{fit}
\end{equation}
to fit the rapidity distribution of the correlation, where $F_1(\Delta\eta)=A_1 e^{-\Delta\eta^2/\sigma_1^2}$ is the DF-wake valley, $F_2(\Delta\eta,\eta_j)=A_2e^{-(\Delta\eta+\eta_j)^2/\sigma_2^2}$ is the MPI ridge, $F_3(\eta_j)$ is the self-normalized Gaussian-like rapidity distribution of $\gamma$-triggered jets from CoLBT-hydro simulations, and $\eta_{j1,j2}$ define the jet rapidity range in the analysis. We assume that the DF-wake valley and MPI ridge are both Gaussian-like. The dot-dashed line in Fig.~\ref{fig:etacorr}(a) demonstrates the robustness of the 2-Gaussian fit to the double peak structure. We will only focus on soft hadrons in the $p_{\rm T}=$1--2 GeV/$c$ range in the following analyses.

\begin{figure}
\centerline{\includegraphics[scale=0.5]{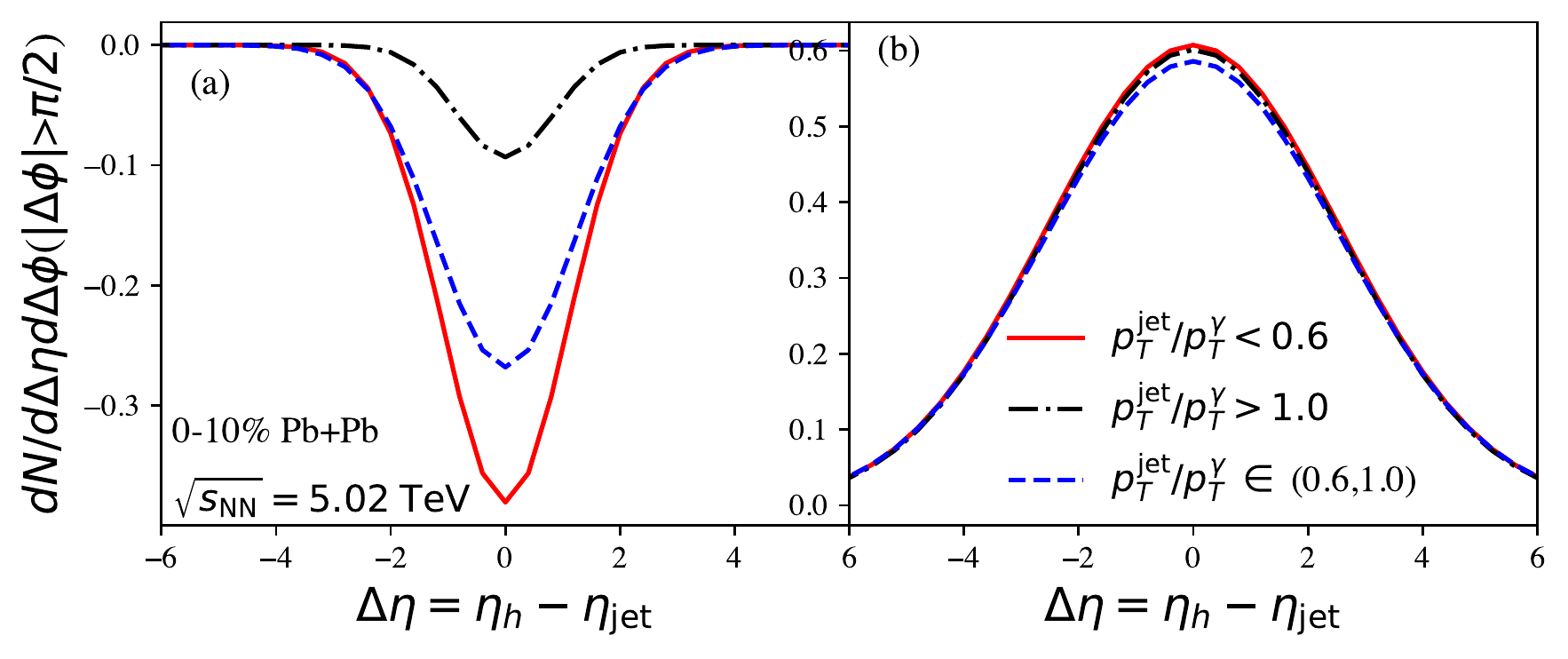}}
\caption{(a) Diffusion wake valley and (b) MPI ridge in $\gamma$-triggered jet-hadron correlation in $|\Delta\phi|>\pi/2$ as a function of $\Delta\eta$ with different ranges of $x_{j\gamma}=p_{\rm T}^{\rm jet}/p_{\rm T}^\gamma$ in 0-10\% central Pb+Pb collisions at $\sqrt{s_{\rm NN}}=5.02$ TeV from CoLBT-hydro. The $\gamma$-jet kinematics is the same as in Fig.~\ref{fig:asym}.}
\label{fig:asyeta}
\end{figure}

Since the $\gamma$-jet asymmetry distribution in Pb+Pb collisions is modified by jet energy loss as shown in Fig.~\ref{fig:asym}, one can use $x_{j\gamma}$ to select events with different jet energy loss which is controlled by the jet propagation length for a given strength of jet-medium interaction.
Shown in Fig.~\ref{fig:asyeta} are the rapidity distributions of (a) the DF-wake valley and (b) MPI ridge from the 2-Gaussian fit to the jet-hadron correlation in 0-10\% central Pb+Pb collisions at $\sqrt{s_{\rm NN}}=5.02$ TeV in the azimuthal angle region $|\Delta\phi|>\pi/2$ opposite to the jet direction for $x_{j\gamma}<0.6$ (red solid), $x_{j\gamma}\in(0.6,1.0)$ (blue dashed) and $x_{j\gamma}>1$ (black dot-dashed). One can see, in events with smaller values of $x_{j\gamma}$ the DF-wake valley is deeper because of  longer propagation length and larger jet energy loss. The MPI ridge, on the other hand, has a very weak and nonmonotonic dependence on $x_{j\gamma}$ due to the nonmonotonic dependence of the propagation length on $x_{j\gamma}$ for minijets from MPI. The bulk background which has to be subtracted in experimental analysis should be independent of the $\gamma$-jet asymmetry.

\noindent{\em \color{blue} 4. Sensitivity to shear viscosity and EoS.---}The spatial structure of the jet-induced Mach cone is known to be sensitive to medium properties such as the EOS and shear viscosity. It is therefore important to check how the DF-wake valley in jet-hadron correlation is sensitive to these medium properties. When using different EoS or shear viscosity in this study, we adjust the overall normalization for the entropy density in the initial conditions for CLVisc such that the final charged hadron rapidity density remains the same as given by the experimental data.

\begin{figure}
\centerline{\includegraphics[scale=0.5]{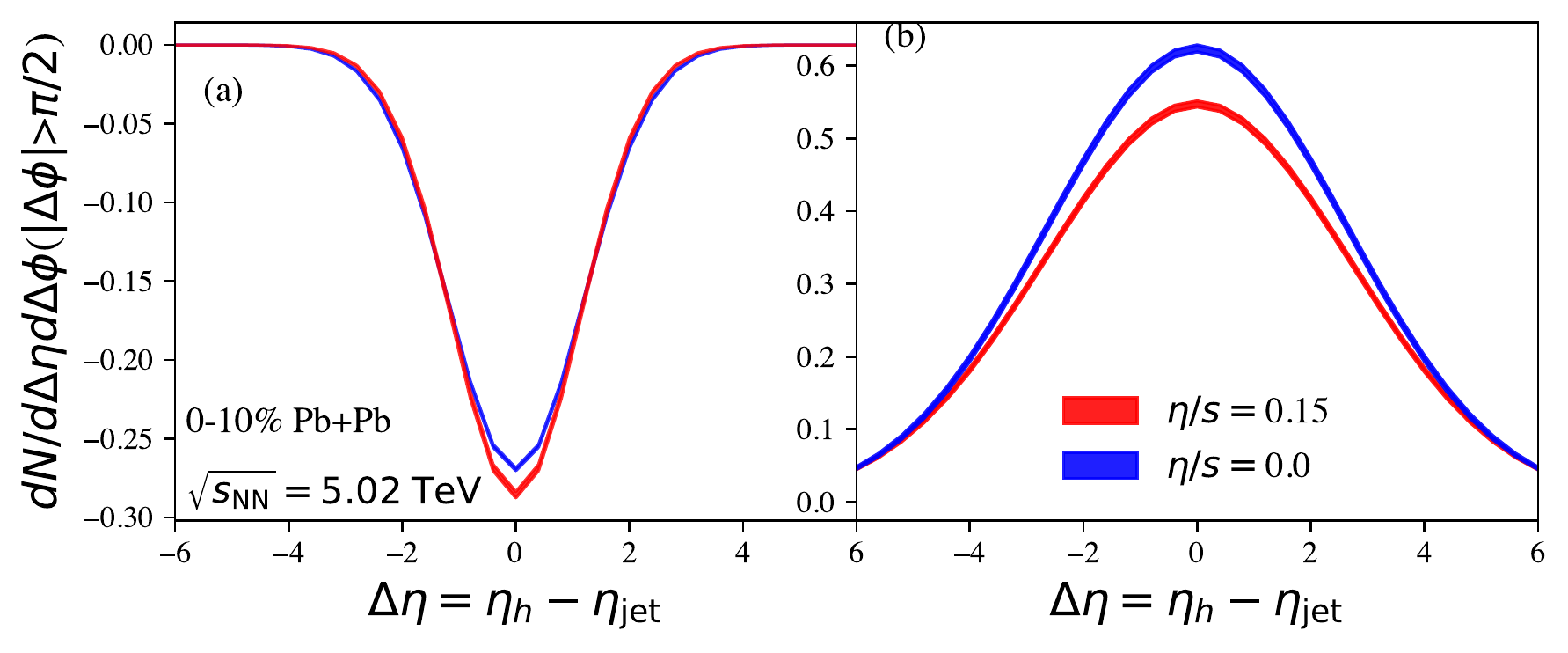}}
\caption{The same as Fig.~\ref{fig:asyeta} except with different values of specific shear viscosity $\eta/s$ in CoLBT-hydro. Bands are numerical errors.}
\label{fig:corr-visc}
\end{figure}

To study the sensitivity to the shear viscosity, we carry out CoLBT-hydro simulations of the same $\gamma$-jet events with two different values of $\eta/s= 0.0$ and 0.15 in CLVisc. The corresponding rapidity distributions of the DF-wake valley and MPI ridge in $|\Delta\phi>|\pi/2$ are shown in Fig.~\ref{fig:corr-visc}.  Shear viscosity is known to increase the transverse flow velocity of the bulk medium and thus increase the slope of the hadron $p_{\rm T}$ spectra. This will suppress the MPI ridge and reduce the DF-wake valley of soft hadrons. In the meantime, the negative shear correction of the longitudinal pressure in the energy-momentum tensor will impede the longitudinal expansion \cite{Song:2007ux}. This will increase the MPI ridge and deepen the DF-wake valley in rapidity. The competition of these two effects leads to a slightly smaller MPI ridge and a deeper DF-wake valley in viscous hydro than in an ideal hydro according to CoLBT-hydro simulations shown in Fig.~\ref{fig:corr-visc}. The statistical errors of the CoLBT-hydro calculations as indicate in the plot are negligible. PYTHIA8 (tune 1) is known to underestimate the MPI multiplicity in $p+p$ collisions by about 10\% \cite{CDF:2004jod,ATLAS:2010kmf,CMS:2010rux,ALICE:2011ac,STAR:2019cie} which should also appear in the MPI ridge in CoLBT-hydro calculations. In principle, the MPI in $\gamma$-jet events can be similarly measured as in $p+p$ collisions or with the mixed-event method as proposed in Ref.~\cite{Yang:2021qtl}.


\begin{figure}
\vspace{0.02in} 
\centerline{\includegraphics[scale=0.5]{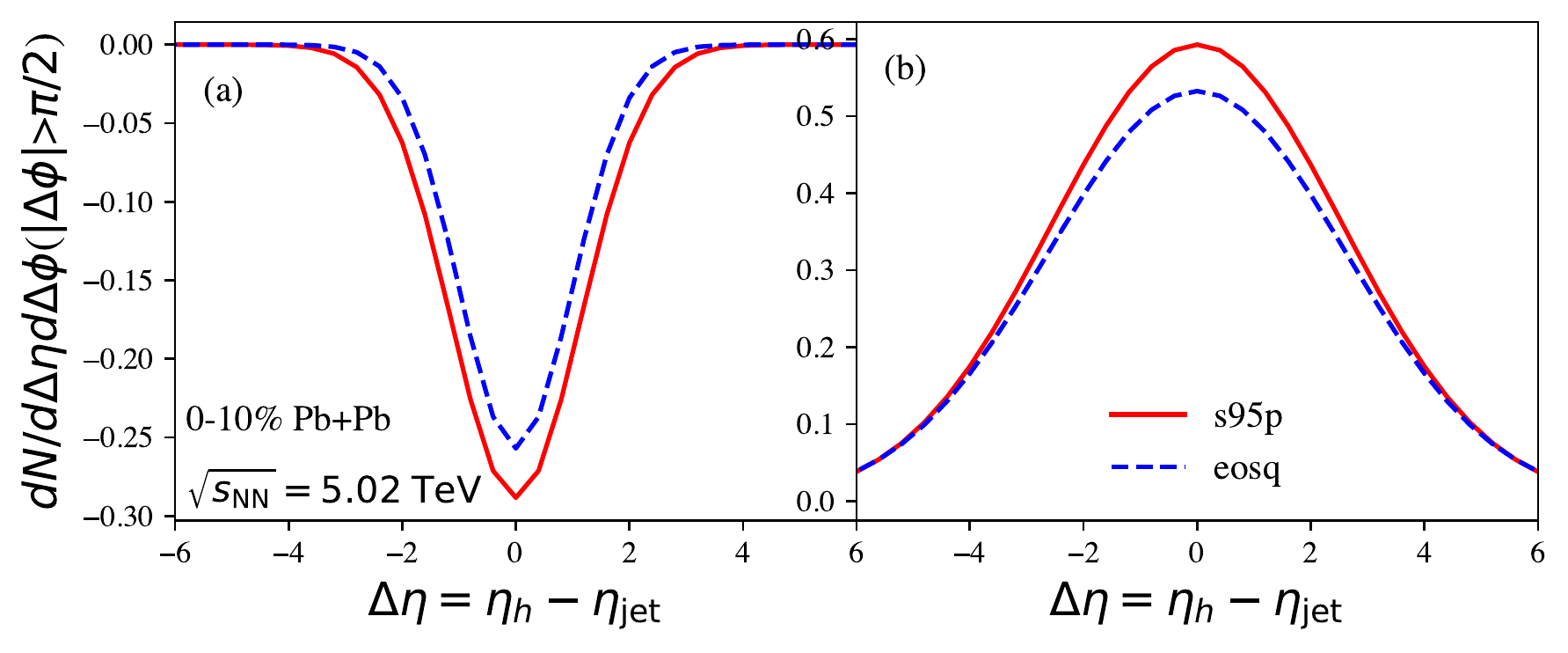}}
\caption{The same as Fig.~\ref{fig:asyeta} except with two different EOS: $s95p$ (solid) and $eosq$ (dashed) in CoLBT-hydro.}
\label{fig:EoS}
\end{figure}

Finally, to check the sensitivity of the medium response to EOS, we consider an EOS ($eosq$) with a first-order phase transition instead of the default EOS ($s95p-v1$) with a rapid crossover \cite{Huovinen:2009yb} in CLVisc. As shown in Fig.~\ref{fig:EoS}, the DF-wake valley is shallower and MPI ridge is smaller in the case of EOS $eosq$ as compared to $s95p$. This can be understood as a consequence of the higher effective sound velocity in $eosq$ than $s95p$ EOS. Since the DF-wake spreads between the Mach cone behind the wave front, a higher sound velocity leads to a larger Mach cone angle and therefore a shallower DF-wake valley. In the meantime, a higher sound velocity will also lead to a stronger radial flow that reduces the soft hadron yield from MPI and the DF-wake valley similarly as one increases the shear viscosity. Distributions of energy density and flow velocity of the medium response with different EOS and shear viscosity can be found in the supplemental material as an illustration.

\noindent{\em \color{blue} 5. Summary.---}We have explored the 3D structure of the diffusion wake induced by $\gamma$-triggered jets in Pb+Pb collisions at the LHC energy within the CoLBT-hydro model. We found that the 2D jet-hadron correlation in azimuthal angle and rapidity has a valley structure in the opposite direction of the jet due to the diffusion wake on top of a MPI ridge along the azimuthal angle. This unambiguously structure in $\gamma/Z$-jet events should be measurable in experiments at RHIC and LHC \cite{CMS:2021otx}. Using a 2-Gaussian fit, we extract the diffusion wake valley and MPI ridge in the jet-hadron correlation. The diffusion wake valley is the deepest in the opposition direction of the jet and increases with the jet energy loss as characterized by the $\gamma$-jet asymmetry. Its sensitivities to the shear viscosity and EOS are modest after constraints on the bulk spectra are taken into account. Nevertheless, future experimental data on the diffusion wake together with other observables can provide combined constraints on the viscosity and EOS of QGP.

\noindent{\em \color{blue} Acknowledgement.---}
We thank Yayun He and Chi Ding for helpful discussions. This work is supported in part by NSFC under Grants No. 11935007, No. 11861131009, No. 11890714 and No. 12075098,
by Fundamental Research Funds for Central Universities in China, by DOE under  Contract No. DE-AC02-05CH11231, by NSF under Grant No. ACI-1550228 within the JETSCAPE and OAC-2004571 within the X-SCAPE Collaboration, by EU ERDF and H2020 Grant 82409, ERC Grant ERC-2018-ADG-835105, Spanish AEI Grant FPA2017-83814-P and MDM- 2016-0692, Xunta de Galicia Research Center accreditation 2019-2022. Computations in this study are performed at the NSC3/CCNU and NERSC.

\appendix

\section{Supplemental material}

\begin{figure}[h]
\vspace{-0.30in}
\centerline{\includegraphics[scale=0.35]{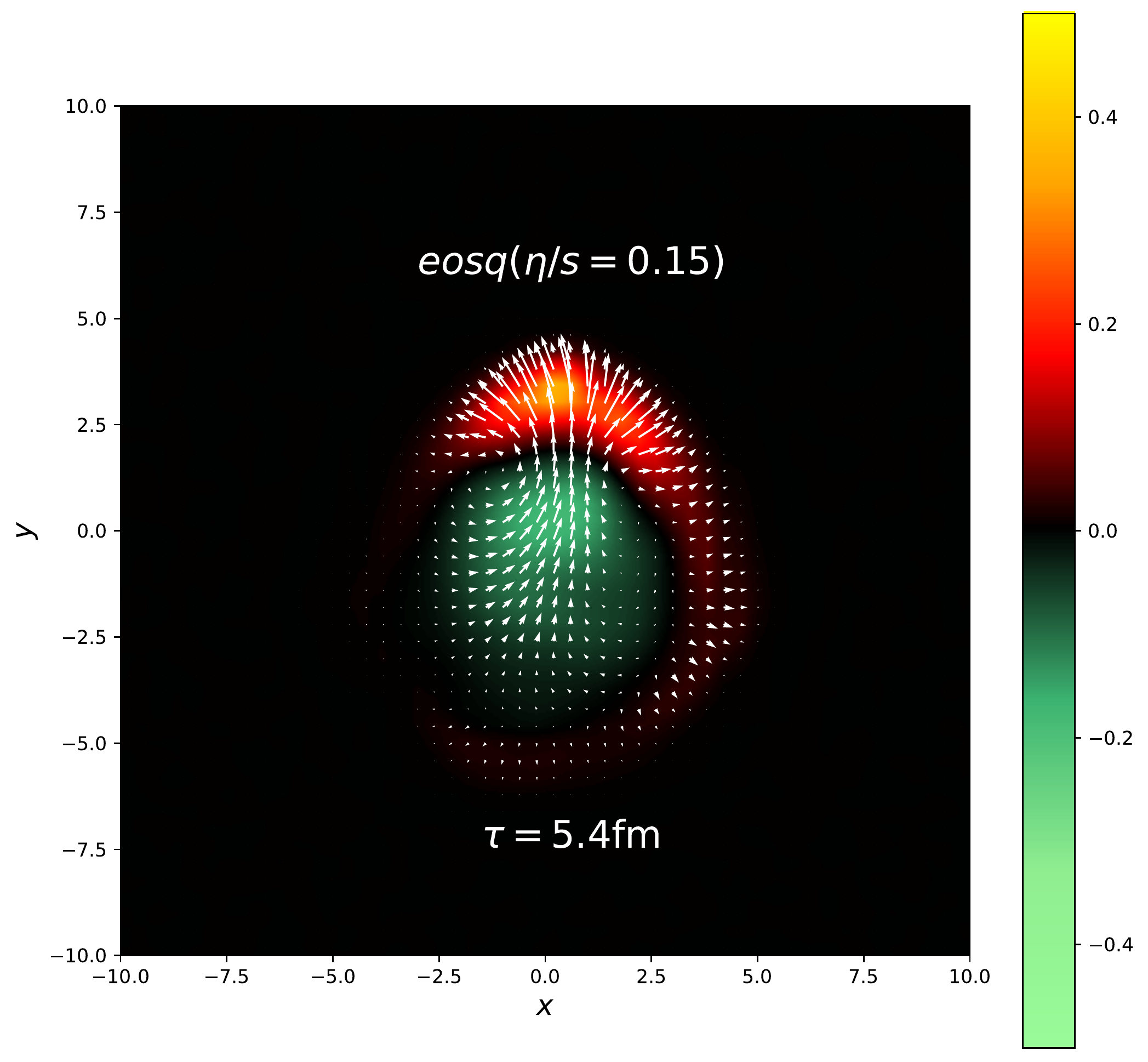}}
\vspace{-0.55in}
\centerline{\includegraphics[scale=0.35]{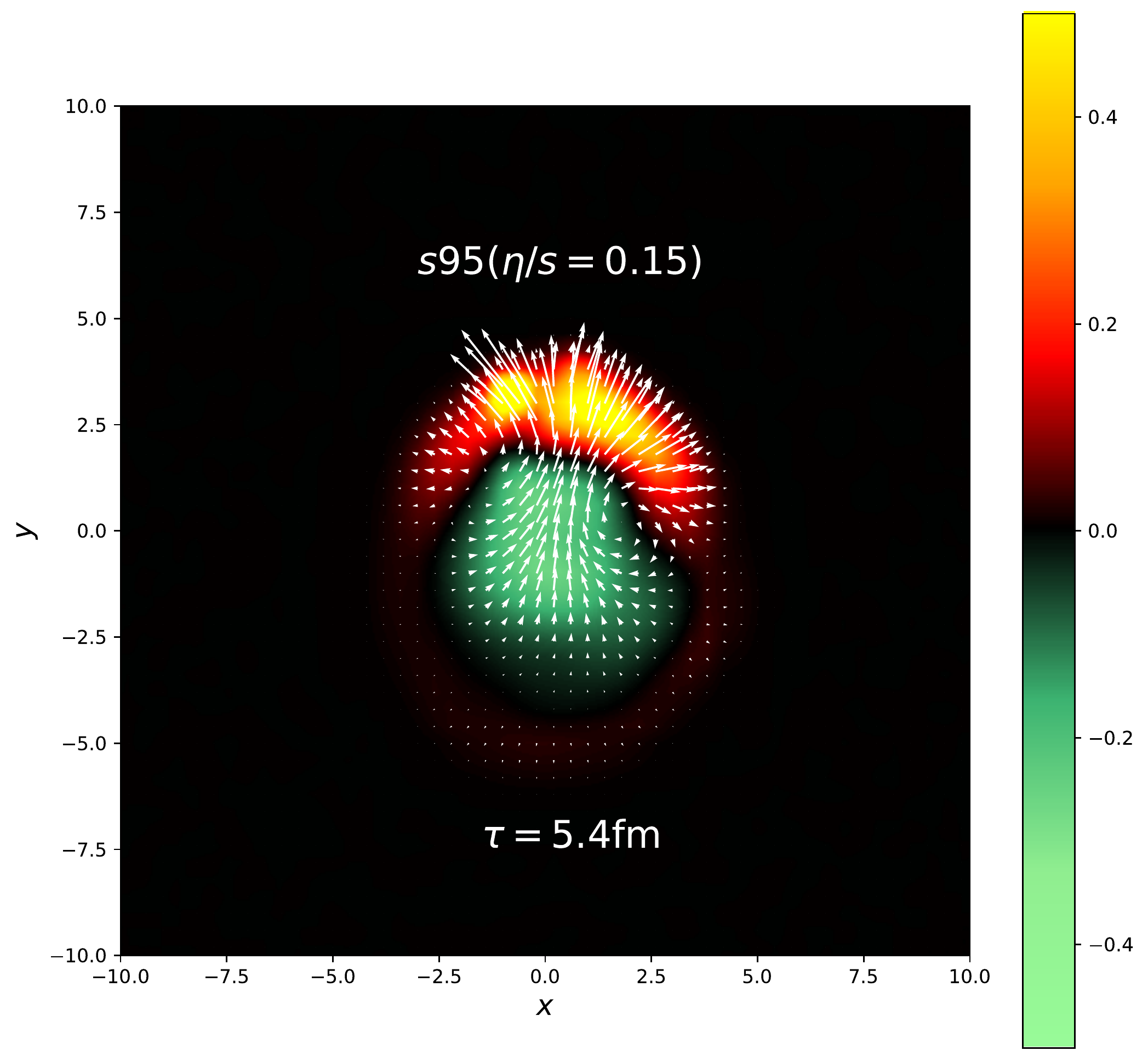}}
\vspace{-0.55in}
\centerline{\includegraphics[scale=0.35]{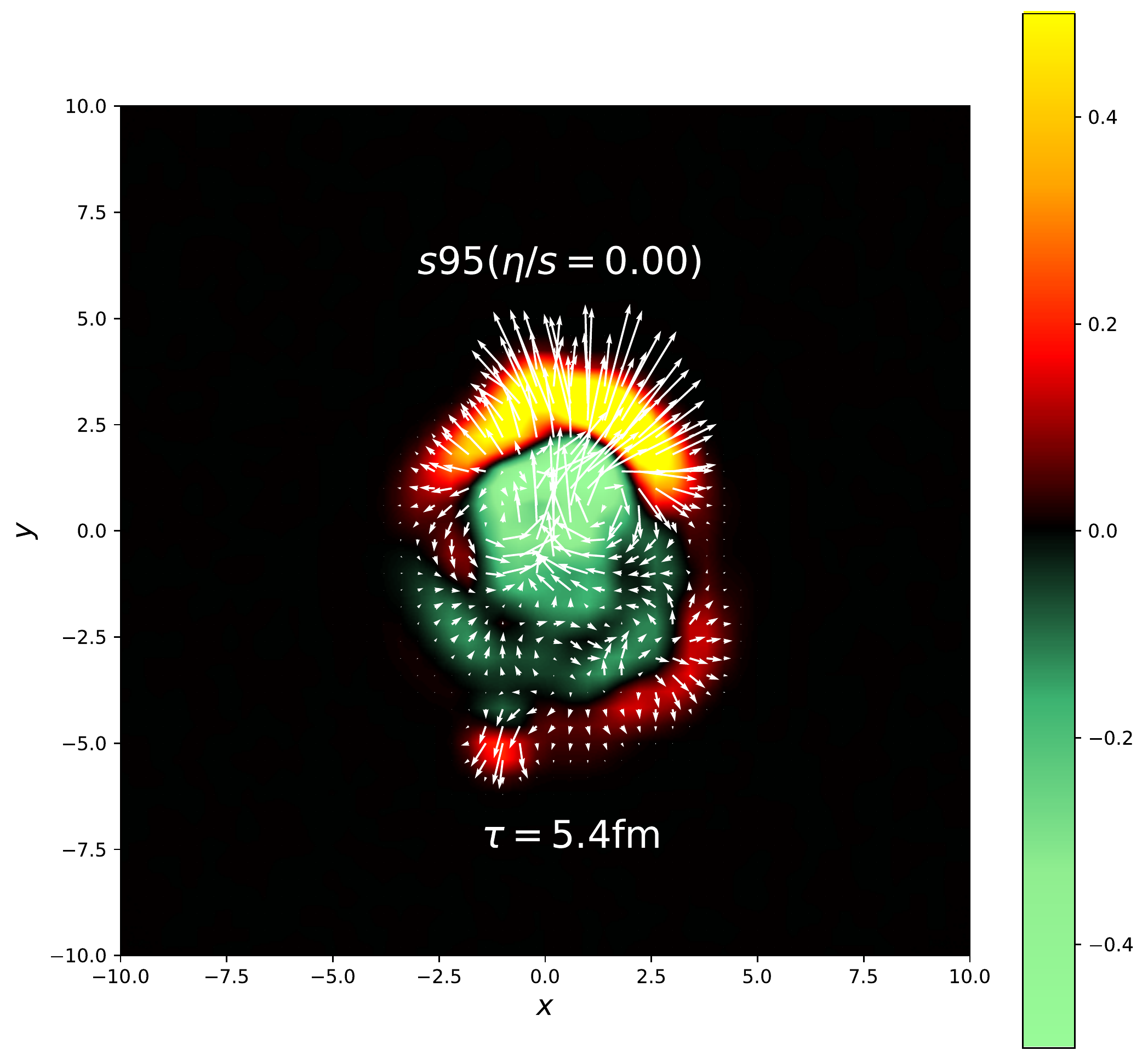}}
\caption{The energy density (color scale) and flow velocity (arrows) distribution of the medium response  in $x$-$y$ plane at $\tau$=5.4 fm/$c$ induced by a $\gamma$-jet with $p_T^\gamma=60$ GeV/$c$ in a 0-10\% central Pb+Pb event at $\sqrt{s}=5.02$ TeV. The $\gamma$-triggered jet is initially produced at $(x,y)=(0,-1.0)$ traveling in the $y$ direction. The EOS and value of $\eta/s$ used in the hydro are indicated in each panel.}
\label{fig:qui_eos}
\end{figure}

\begin{figure}[h]
\centerline{\includegraphics[scale=0.35]{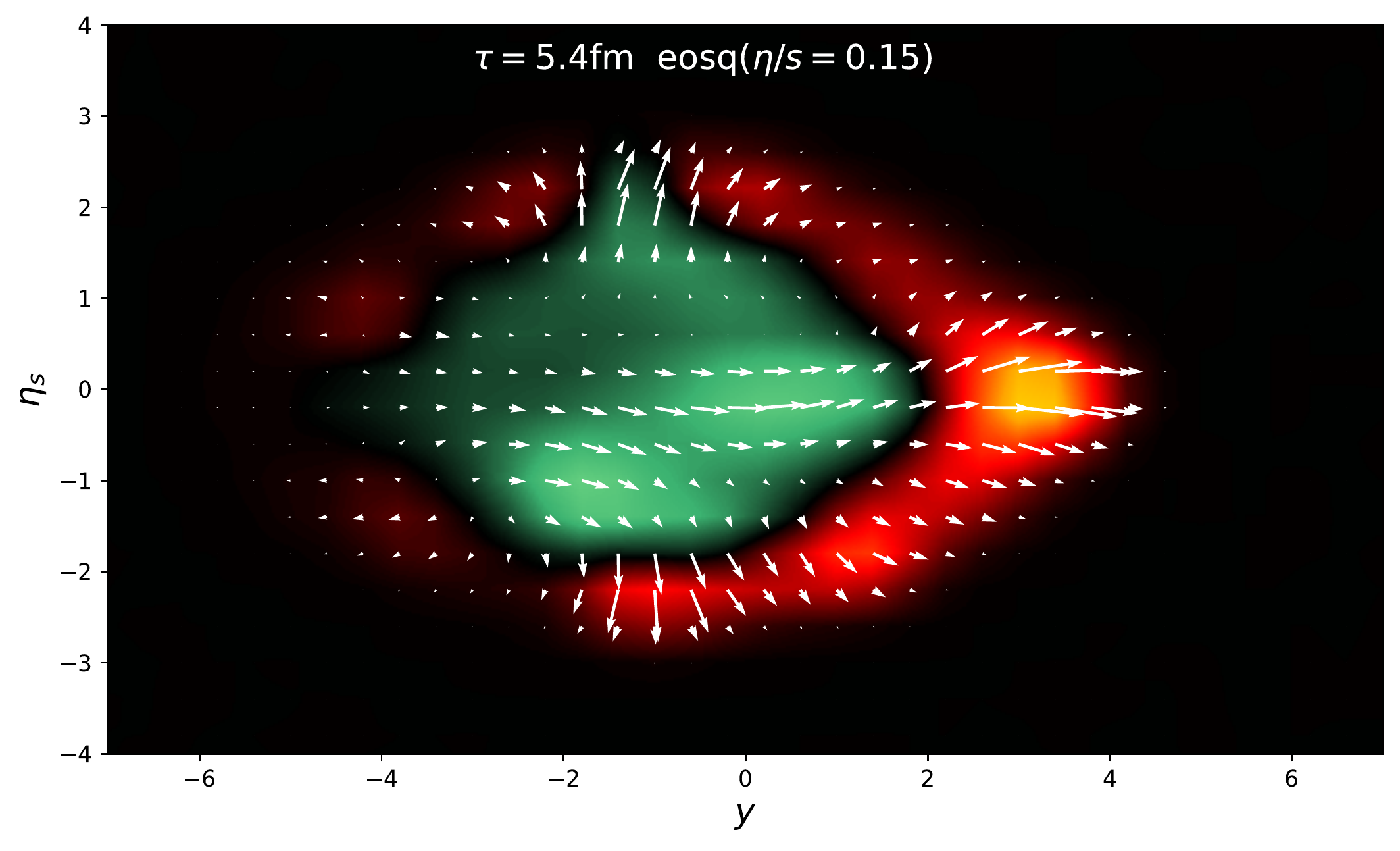}}
\vspace{-0.26in}
\centerline{\includegraphics[scale=0.35]{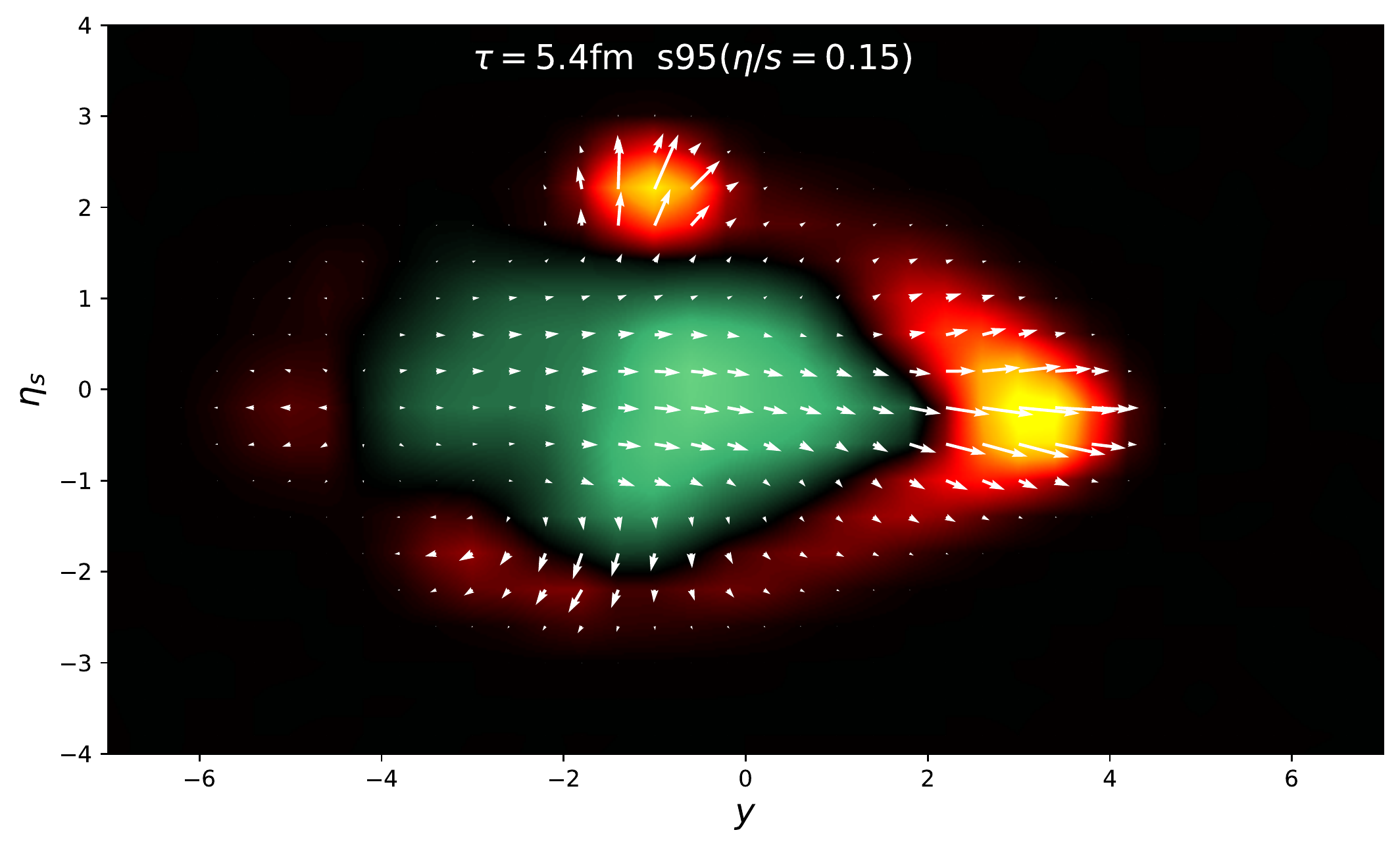}}
\vspace{-0.26in}
\centerline{\includegraphics[scale=0.35]{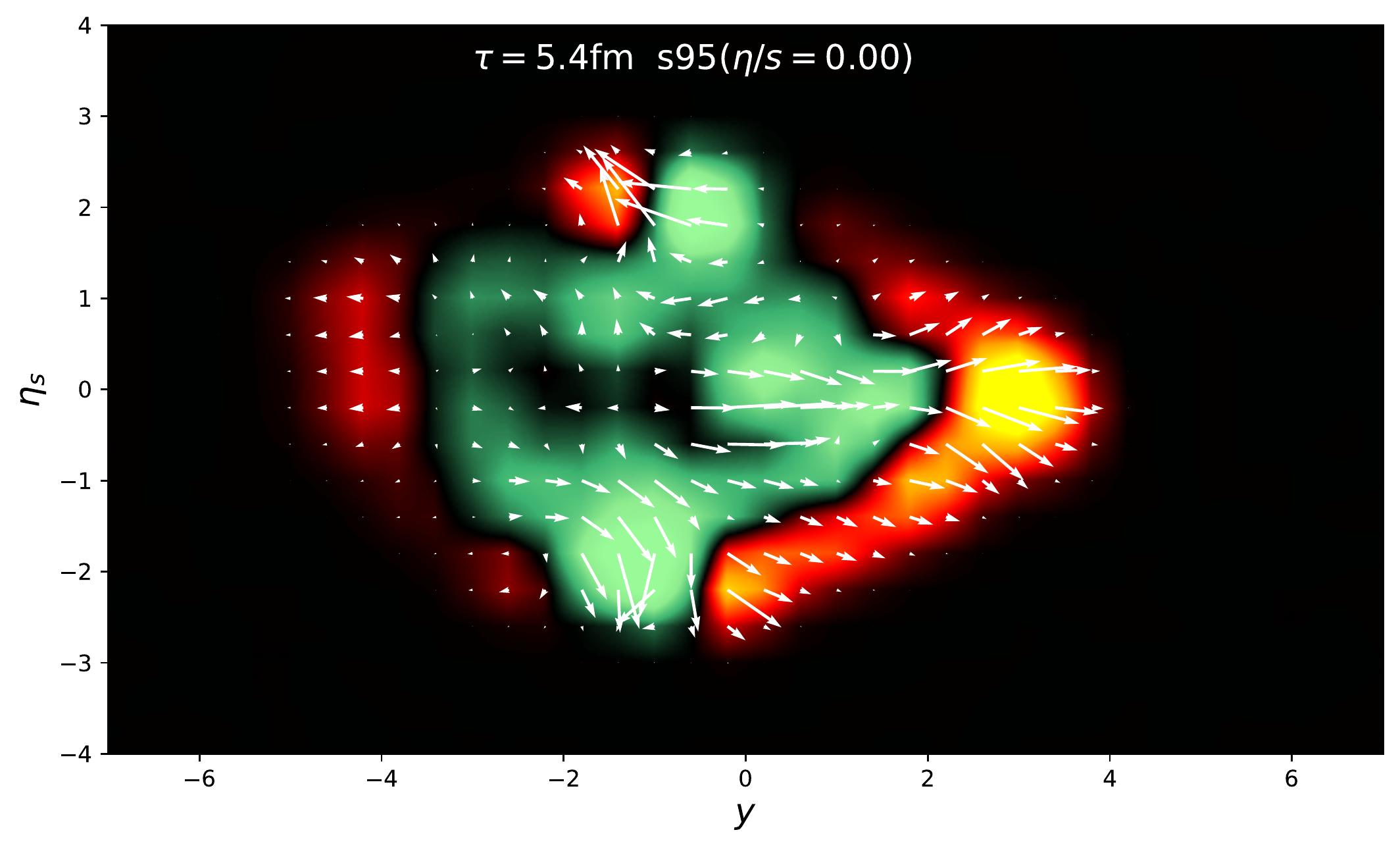}}
\caption{The same as Fig.~\ref{fig:qui_eos} except in the $y$-$\eta_s$ plane.}
\label{fig:ed_eos2}
\end{figure}

In this supplemental material, we provide quiver plots (energy density and flow velocity) both in the $x$-$y$ transverse plane (Fig.~\ref{fig:qui_eos}) and $y$-$\eta_s$ plane (Fig.~\ref{fig:ed_eos2}) of the medium response at $\tau=5.4$ fm/$c$ induced by a $\gamma$-triggered jet with $p_T^\gamma=60$ GeV/$c$ in a 0-10\% central Pb+Pb event for different shear viscosity and equations of state (EOS). These jet-induced medium responses are the differences between hydrodynamic evolution of the bulk medium with and without the $\gamma$-jet. In all the cases, the diffusion wake (negative energy density) is present which leads to the diffusion-wake valley in the final jet-hadron correlation in the momentum-space. 

The finite value of shear viscosity is clearly seen to smooth out both the energy density and flow velocity fluctuations as compared to the case of an ideal fluid ($\eta/s=0.0$) when one compares the middle to lower panels of Figs.~\ref{fig:qui_eos} and \ref{fig:ed_eos2}. The expansion in the longitudinal direction for the ideal fluid (lower panel of Fig.~\ref{fig:ed_eos2}) is clearly reduced due to the negative shear correction to the longitudinal pressure when shear viscosity is finite (middle panel of Fig.~\ref{fig:ed_eos2}).  

The negative energy density due to the diffusion wake in the case of the $s95p$ EOS with a rapid cross-over (middle panels of Figs.~\ref{fig:qui_eos} and \ref{fig:ed_eos2}) is indeed deeper than that for $eosq$ with a first-order (upper panels of Figs.~\ref{fig:qui_eos} and \ref{fig:ed_eos2}) as we have argued in the paper. This leads to a deeper DF-wake valley in the $\gamma$-hadron correlation for the case of $s95p$ EOS.

\begin{thebibliography}{99}
\bibitem{footnote}
 The finite spatial grid size and time step in CLVisc hydro lead to numerical errors in the final hadron spectra from the medium response due to the discretized freeze-out hyper-surface. We have estimated these errors by imposing energy-momentum conservation and corrected the errors by dividing a correction factor  which is 1.20 for $\eta/s=0.15$ and 1.35 for $\eta/s=0$ with grid size $dx=dy=0.16$ fm, $dz=0.2$ fm, time step $dt=0.02$ fm/$c$ and the numbers of grids for the calculation of hyper surface are $n_x=n_y=n_z=2$ and $n_t=20$.
\end{thebibliography}
\end{document}